\documentclass{svproc}
\usepackage{url}

\usepackage{graphicx}

\usepackage{array}
\newcolumntype{C}[1]{>{\centering\let\newline\\\arraybackslash\hspace{0pt}}m{#1}}

\usepackage{color}
\definecolor{new_green}{rgb}{0,0.5,0}
\definecolor{forestgreen}{RGB}{34,139,34}
\definecolor{new_color}{RGB}{255,153,51}

\begin{document}
\mainmatter              
\title{Fast Flux Service Network Detection via Data Mining on Passive DNS Traffic}
\titlerunning{Fast Flux Detection via Data Mining on Passive DNS Traffic}
%
\author{Pierangelo Lombardo \inst{1} \and Salvatore Saeli \inst{1} \and Federica Bisio \inst{1} \and Davide Bernardi \inst{1} \and Danilo Massa \inst{1}}
\authorrunning{Lombardo, P., Saeli, S., et al.} 
\institute{$^1\,$aizoOn Technology Consulting, Strada del Lionetto 6, 10146 Turin, Italy
\email{[name].[surname]@aizoongroup.com}
}

\maketitle              
\begin{abstract}
In the last decade, the use of fast flux technique has become established as a common practice to organise botnets in Fast Flux Service Networks (FFSNs), which are platforms able to sustain illegal online services with very high availability. 
In this paper, we report on an effective fast flux detection algorithm based on the passive analysis of the Domain Name System (DNS) traffic of a corporate network.
The proposed method is based on the near-real-time identification of different metrics that measure a wide range of fast flux key features; the metrics are combined via a simple but effective mathematical and data mining approach.
The proposed solution has been evaluated in a one-month experiment over an enterprise network, with the injection of 
pcaps associated with different malware campaigns, that leverage FFSNs and cover a wide variety of attack scenarios.
An in-depth analysis of a list of fast flux domains confirmed the reliability of the metrics used in the proposed algorithm and allowed for the identification of many IPs that turned out to be part of two notorious  FFSNs, namely Dark Cloud and SandiFlux, 
to the description of which we therefore contribute.
All the fast flux domains were detected with a very low false positive rate; a comparison of performance indicators with previous works show a remarkable improvement.
\end{abstract}

\section{Introduction}\label{sec:intro}

During the last few years, the number of cyberattacks with relevant financial impact and media coverage has been constantly growing. As a result, many companies and organizations have been reinforcing investment to protect their networks, with a resultant increase in the research on this topic \cite{acs}.

Over the last two decades, botnets have represented one of the most prominent sources of threats on the internet: they are networks of compromised computers (popularly referred to as zombies or bots), which are controlled by a remote attacker (bot herder).
Botnets provide the bot herder with massive resources (bandwidth, storage, processing power), allowing for the implementation of a wide range of malicious and illegal activities, like spam, distributed denial-of-service attacks,  spreading of malware (such as ransomware, exploit kits, banking trojans, etc.) \cite{alieyan2017survey,berger2016mining,bisio2017real,chahal2016tempr,holz2008measuring}.

A common practice for bot herders is to organise their bots in Fast Flux Service Networks (FFSNs): some bots, chosen from a pool of controlled machines, are used as front-end proxies that relay data between a (possibly unaware) user and a protected hidden server. 
The technique behind these structures is the \emph{fast flux}, i.e., the rapid and repeated changing of an internet host and/or name server resource record in a Domain Name System (DNS) zone, resulting in rapid changes of the IP addresses to which the domain resolves.
FFSNs make the tracing and the recovery of all the infected components extremely difficult, thus allowing for a very high availability for illegal online services  related to phishing, dumps stores, and  distribution of ransomware, info stealers, and click fraud \cite{crowder2016dark,lin2013genetic,nazario2008net,salusky2007know,soltanaghaei2015detection,zhou2015survey}.

FFSNs have been known to cybersecurity experts for more than one decade \cite{holz2008measuring,salusky2007know}, but in the last few years it has been obtaining a spotlight
\cite{almomani2018fast,berger2016mining,chahal2016tempr,jiang2017exploring,ruohonen2017investigating,stevanovic2017method}.
The renewed interest is related to the studies of large botnets (e.g., Dark Cloud, also known as Zbot network, and the most recent SandiFlux) which make massive usage of fast flux \cite{sandiflux,crowder2016dark,katz2017digging}.
The standard approach to FFSNs detection is via the so-called \emph{active} DNS analysis, i.e., by actively querying some domains and by collecting and analysing the answers: this strategy has been widely explored and allows for extensive analyses of botnets \cite{almomani2018fast,holz2008measuring,hsu2010fast,jiang2017exploring,katz2017digging,lin2013genetic,martinez2013real,nazario2008net,passerini2008fluxor}.

Instead, the algorithm described in the present work relies on \emph{passive} analysis of the DNS traffic of a single network: it detects the fast flux domains without interaction with the network traffic, thus making the algorithm completely transparent inside and outside the monitored network; in particular, it cannot be uncovered by the attackers, who often control the authoritative name servers responsible for responding to DNS queries about their fast flux domains \cite{perdisci2012early}.
The proposed detection approach has been evaluated in a 30-day-long experimental session over the network described in Sect.~\ref{sec:experimental}. The performance is much higher compared to a state-of-the-art analogous method \cite{soltanaghaei2015detection}.  
Moreover, the analysis was performed near-real-time: the average execution time of the algorithm was 25 seconds, while the average time between two subsequent runs of the algorithm was 3 minutes (see Sect.~\ref{sec:aramis} for more details), meaning that the average detection time for fast flux domains was less than 2 minutes. 

As an additional test of the proposed approach, we examined the IPs --- collected via active DNS analysis --- associated with a list of fast flux domains gathered from 
 \cite{sandbox1,sandbox3,sandbox2,virustotal}. This investigation confirmed the reliability of the metrics used in the fast flux detection method proposed herein
 and allowed for the identification of more than $10000$ IPs, some of which are likely 
 associated with compromised hosts, which turned out to be part of two notorious botnets, namely Dark Cloud and SandiFlux, to the description of which we therefore contribute.

The paper is structured as follows. In Sect.~\ref{sec:background}, we 
discuss the most relevant features of FFSNs, with an outline of related works.
In Sect.~\ref{sec:aramis}, we briefly describe \emph{aramis}, the monitoring platform that contains the fast flux detection method which is the focus of this paper and which is described thoroughly in Sect.~\ref{sec:detection}.
Section \ref{sec:experimental} comprises a detailed discussion of the experimental results of the test of the proposed algorithm, while Sect.~\ref{sec:botnet}
contains further investigations on the FFSNs underlying some fast flux domains.
Finally, we discuss  possible future developments in Sect.~\ref{sec:conclusions}.

\section{Background and Related Work}\label{sec:background}

One of the first works providing an overview of the fast flux attacks was the Honeynet project \cite{salusky2007know}. In order to explain hidden operations executed by botnets, authors gave examples of both single and double fast flux mechanisms: while the first rapidly changes the A records of domains, the latter frequently changes both the A records and the NS records of a domain. The interested reader can find a review and a classification of fast flux attacks in \cite{zhou2015survey}.

Content Delivery Network (CDN) and Round-Robin DNS (RRDNS) are legitimate techniques which are used by large websites to distribute the load of incoming requests to several servers. The response to a DNS query is evaluated by an algorithm which chooses a pool of IPs from a large list of available servers whose number can be of the order of thousands (see Sect.~\ref{sec:botnet} for some examples). As a result, the  behaviour in terms of DNS traffic is very similar to the one of a FFSN, and indeed CDNs and RRDNSs represent the typical false positives in  fast flux detection algorithms \cite{holz2008measuring,lin2013genetic,soltanaghaei2015detection}.

A large number of approaches have been proposed to detect FFSNs and to distinguish them from legitimate CDNs and RRDNSs. Most of them rely on active DNS analysis, which allows for the collection of a large number of IPs associated with a domain, thus simplifying the FFSNs detection, but they require the resolutions of domains that may be associated with malicious activities \cite{chahal2016tempr,holz2008measuring,jiang2017exploring,lin2013genetic,nazario2008net}. These methods, despite being appropriate for a deep analysis of FFSNs,  have relevant drawbacks in implementations oriented to the monitoring of corporate networks \cite{perdisci2012early,soltanaghaei2015detection}.

Some FFSN detection methods based on passive DNS analysis have been proposed. 
Some of them analyse the DNS traffic of a whole Internet Service Provider (ISP), thus taking in input the DNS traffic generated by many  different networks.
Perdisci et al.~\cite{perdisci2012early}, in particular, performed a large-scale passive analysis of DNS traffic. They extract some relevant features from the DNS traffic and classified the domains via a C4.5 decision tree classifier.
Berger et al.~\cite{berger2016mining} and 
Stevanovic et al.~\cite{stevanovic2017method} proposed two other approaches to analyse the DNS traffic of an ISP. Both methods are based on a tool called DNSMap and  classify the bipartite graphs formed by the collected fully qualified domain names and the associated IPs. The first method searches for generic malicious usage of DNS, while the latter focuses on FFSNs.

Soltanaghaei and Kharrazi \cite{soltanaghaei2015detection}, finally, proposed a method for passive DNS analysis of a network which requires a history for each domain to be evaluated  and achieved 94.44\% detection rate and 0.001\% false positive rate in their best experiment.
Our algorithm employs a similar approach, but, with a more careful choice of the metrics achieves better results, while performing a near-real-time analysis (see Sects.~\ref{sec:detection} and \ref{sec:experimental} for details).

\section{aramis}\label{sec:aramis}

The proposed fast flux 
detection technique is included in a commercially available network security monitoring platform called \emph{aramis} (Aizoon Research for Advanced Malware Identification System) \cite{aramis,bisio2017real}. This software automatically identifies different types of malware and attacks in near-real-time, it is provided with dedicated hardware%
\footnote{E5-2690 2.9GHz x 2 (2 sockets x 16 cores)
16 x 8GB RAM, 1.1TB HDD}%
, and its structure can be outlined in four phases:

\begin{enumerate}
\item Collection: sensors located in different nodes of the network gather data, preprocess them in real-time and send the results to a NoSQL database.
\item Enrichment: data are enriched in the NoSQL database using the information obtained from the aramis Cloud Service, which collects intelligence from various open-source intelligence sources and from internally managed sources.
\item Analysis: stored data are processed by means of two types of analysis: (i) advanced cybersecurity analytics which highlight specific patterns of attacks, among which Domain Generation Algorithms (DGAs) \cite{bisio2017real} and fast flux, and (ii) a machine learning engine which spots deviations from the usual behaviour of each node of the network.
\item Visualization: results are presented in dashboards to highlight anomalies.
\end{enumerate}

The cycle of the four phases restarts after a time $\Delta t$ which slightly depends on the traffic flow analysed and amounts to $182\pm 36$\,seconds on the network described in Sect.~\ref{sec:experimental}. A time $\Delta t$ of this magnitude 
is the best trade-off between the near-real-time requirement and the need of a large amount of data in order to have statistically significant results.

\section{Detection Method}\label{sec:detection}

The aim of the proposed detection method is the near-real-time identification of malicious fast flux via the passive monitoring of the DNS traffic of a single network.
To this purpose, the method is composed of three steps of analysis.
(i) Filtering: queries which are known to be non-malicious (e.g., popular domains, known CDNs, local domains, etc.) are removed.
(ii) Metrics identification: some key indicators are calculated over the queries remaining after filters.
(iii) Identification: the metrics are used to identify malicious fast flux among the queries.

The three steps have been constructed by combining information on the FFSNs --- acquired from the literature --- with a simple but effective mathematical and data mining approach. The parameters of the model have been estimated over a \emph{validation set} formed by 30-days of DNS traffic captured from the network described in Table~\ref{tab:network_val}, and by 12 pcaps associated with different malware campaigns that leverage FFSNs, collected from the public repository \cite{maltrafficrepo}.

%
%
\begin{table}[!ht]
\centering
\advance\leftskip-2mm
\caption{Validation-network description}\label{tab:network_val}
\begin{tabular}{ | p{7.1cm} | c | c |}
  \hline
  &\,\textbf{30-days total}\,&\,\textbf{one-hour average}\,\\
  \hline
   \textbf{N. of machines} & 261 & - \\
  \hline
  \textbf{N. of connections} & 80\,M & 111\,k\\
  \hline
  \textbf{N. of resolved A-type DNS queries} & 12\,M  & 17\,k \\
  \hline
  \textbf{N. of unique resolved A-type DNS queries} & 381\,k& 527\\
  \hline
\end{tabular}
\end{table}
%
%

\subsection{Filters}\label{sec:filters}

The algorithm receives resolved DNS requests of type A (which return 32-bits IPv4 addresses, in accordance with \cite{rfc1035}) collected near-real-time  from the monitored network. 
The first step consists in the application of the filters reported in Table~\ref{tab:filters} to the retrieved queries.

%
%
\begin{table*}[!ht]
\centering
\begin{minipage}{\textwidth}
\caption{Filters description}\label{tab:filters}
\advance\leftskip-2mm
\begin{tabular}{ | C{2.75cm} | C{9.65cm} |}
  \hline
  \textbf{Type} & \textbf{Description} \\
  \hline
  White list domains & Domains known to be trusted, e.g., the ones associated with crypto currencies (if their use is allowed in the network): the underlying peer-to-peer networks are, in many respects, similar to botnets. \\
  \hline
  Popular domains & Top 100 domains collected inside the network under analysis, web URLs of the 500 world biggest companies provided by Forbes \cite{forbes} and top 10000 domains in the world provided by Alexa \cite{alexa}. \\
  \hline
Configuration words & Domains containing certain substrings (e.g., related to network system and structure) represent congenital traffic. \\
\hline
  Overloaded DNS & In order to provide anti-spam or anti-malware techniques, DNS queries are sometimes overloaded, thus causing possible noise. \\
  \hline
  Local and corporate domains & 
  These domains represent a high percentages of the legitimate DNS traffic in a corporate network.\\
  \hline
  CDNs and RRDNSs & These are the most common sources of false positives in fast flux detection algorithms; aramis (see Sect.~\ref{sec:aramis}) includes a function (with a structure similar to that of the proposed fast flux detection algorithm) which periodically updates a white list with the main CDNs and RRDNSs detected in the monitored network, thus allowing for a substantial speed up.\\
  \hline
  Queries with large TTL & According to the literature, malicious fast flux are characterised by a short TTL \cite{holz2008measuring,nazario2008net,soltanaghaei2015detection}, 
therefore queries with a TTL larger than $1800\,{\rm s}$ are filtered. \\
  \hline  
\end{tabular}
\end{minipage}
\end{table*}
%
%

\subsection{Metrics Identification}\label{sec:metrics}

The DNS requests that survive the filters described in Sect.~\ref{sec:filters} are integrated with the history of the previous 30 days, saved locally. This allows for a more accurate evaluation of the behaviour of the domains, however an assessment is already possible when the first answer is received.
Among the remaining domains there are many new emerging CDNs%
\footnote{The filter mentioned in Table~\ref{tab:filters} detects a CDN only when it has a sufficient history.}  
and, in order to distinguish them from the FFSNs
--- which is the main challenge in malicious fast flux detection --- we identified some key indicators.
Some of these indicators can be already evaluated after a single query (we call them \emph{static metrics}), while others need a certain history (\emph{history-based metrics}).
The information regarding Autonomous Systems (ASs) and public networks used in the following metrics  are retrieved from \cite{maxmind}.

\subsubsection{Static Metrics.}

The metrics described in this section are evaluated over all the IPs collected.

\emph{Maximum Answer Length}. A relevant metric for the detection of malicious fast flux is the number of IPs returned in a single A query. In particular, we consider the maximum $m_{\rm al}$ of such value: a malicious fast flux is believed to typically have a $m_{\rm al}$ larger than a legitimate fast flux \cite{holz2008measuring,zhou2015survey}. 

\emph{Cumulative Number of IPs}. Malicious fast flux typically employ a larger number  of IPs ($n_{\rm IP}$) compared with CDNs, due to the lower reliability of each single node \cite{soltanaghaei2015detection}.

\emph{Cumulative Number of Public Networks}. Since the botnet underlying a malicious fast flux contains infected machines which are typically distributed quite randomly in different networks, the same is expected to be true for the IPs retrieved by the related queries \cite{holz2008measuring,zhou2015survey}. For this reason a malicious fast flux typically has a number of public networks ($n_{\rm net}$) larger than a legitimate CDN.

\emph{Cumulative Number of ASs}. For the same reason described above, FFSNs typically have a number of ASs ($n_{\rm AS}$) larger than legitimate CDNs.

\emph{AS-Fraction}.
The analysis of some preliminary fast flux pcaps revealed that, despite being in general very useful, in some cases the absolute number of AS was not a distinctive feature, while its ratio with the number of IPs was more appropriate. For this reason we defined the metric 
 \begin{equation}\label{eq:as-frac1}
 f_{\rm AS}=\frac{n_{\rm AS}-1}{n_{\rm IP}}, 
 \end{equation}
 which quantifies the degree to which the IPs are dispersed in different AS. This quantity takes values from $f_{\rm AS}=0$ (when all the IPs are in the same AS) to $f_{\rm AS}\sim 1$ (when each IP is in a different AS and the number of IPs is large).  
In order to preserve these properties and to encode the additional information about the typical scales associated with $n_{\rm AS}$ for CDNs and FFSNs respectively, we rescaled $f_{\rm AS}$ as described below. 
The first rescaling is%
\footnote{Hereafter, the left and right hand sides of the arrow represent the quantity before and after the rescaling respectively.}
\begin{equation}\label{eq:as-frac2}
x\ \longrightarrow\ \theta(n_{\rm AS}-n_0) \left[1-e^{-\left( \frac{n_{\rm AS}-n_0}{s}\right)^2} \right]x, 
\end{equation}
where $x=f_{\rm AS}$, $\theta(t)$ is the Heaviside step function (i.e., $\theta(t)$ is 1 for positive $t$ and 0 otherwise), $s$ is a scale representing the average number of ASs in a typical CDN and $n_0$ is a threshold for $n_{\rm AS}$ below which the behaviour is not suspicious from the viewpoint of the number of ASs.%
\footnote{We set $s=2.5$ and $n_0=3$; the first is the average $n_{\rm AS}$ for the top 4 largest CDNs detected in the validation set, while the latter is half the minimum of $n_{\rm AS}$ detected for a fast flux in the validation set.}
The rescaling in Eq.~\ref{eq:as-frac2} reduces $f_{\rm AS}$ when its value is comparable with the $n_{\rm AS}$ expected in a CDN.
The second rescaling applies Eq.~\ref{eq:as-frac2} to the quantity $x=1-f_{\rm AS}$ and reduces it (i.e., increases $f_{\rm AS}$) when $n_{\rm AS}$ is comparable with that of a typical FFSNs.
In this case the scale $s$ represents the average number of ASs in a typical malicious fast flux, while $n_0$ is a threshold for $n_{\rm AS}$ below which we do not increase $f_{\rm AS}$.%
\footnote{We set $s=40$ in agreement with Ref.~\cite{zhou2015survey}, which states that a typical FFSN has a set of IPs distributed among 30--60 ASs, and $n_0=5$, which is the maximum number of ASs detected for a CDN in the validation set.}

\emph{IP-Dispersion.} The analysis of the distribution of the retrieved IPs is another way to understand to which degree the structure underlying FFSN is random and chaotic. We transform the set of the $n$ IPs associated with each query into the 
corresponding positions in the 32-bits IPv4 address space $x_1,... x_n$,%
\footnote{To each IP $n_1.n_2.n_3.n_4$ we associated $x=256^3\, n_1 + 256^2\, n_2 + 256\, n_3 + n_4$.}
 and we define
\begin{equation}\label{eq:ipdispesion}
d_{\rm IP}=\frac{1}{l_n}\,{\rm median}(\Delta\vec{x}),
 \end{equation}
where $\Delta \vec{x}=\left\{ x_i-x_{i-1}\right\}_{i=2}^n$, the $\{x_i\}$ have been ordered so that $x_i\geq x_{i-1}$, and $l_n$ is the average distance if the $n$ IPs were uniformly distributed in  the whole public IPs address space. The IP-dispersion takes value from $d_{\rm IP}=1$ (i.e., when the IPs are uniformely distributed) to $d_{\rm IP}=0$ (i.e., when the IPs are clearly subdivided into a few clusters of close addresses).
A similar idea was used by Nazario et al.~\cite{nazario2008net}, who evaluated the average distance among the $\{x_i\}$, but their metric  is more sensitive to outliers and it is not normalised in the interval [0,1], which is crucial to combine it with the other metrics, as described in Sect.~\ref{sec:ff_identification}.
The FFSNs analysis described in Sect.~\ref{sec:botnet} confirmed that the indicator in Eq.~\ref{eq:ipdispesion} is able to catch the key distribution properties of IPs in a FFSN.

\subsubsection{History-Based Metrics.}

The \emph{history} is constructed by subdividing the queries retrieved from the monitored network in subsequent \emph{chunks}: each chunk contains at least 10 queries and spans a time interval of at least one hour; these two conditions are the minimal requirements to make the metric definitions meaningful from a statistical point of view.
The metrics described in this section are evaluated only if it is possible to construct at least two chunks.

\emph{Change in the set of IPs}. It
is a common belief that, while a CDN typically returns IPs taken from a stable IP-pool, a malicious fast flux  employs the available nodes in the FFSN, which often evolves quickly, and therefore its IP-pool changes from time to time \cite{holz2008measuring,zhou2015survey}. We defined a metric which measures in a very simple way the change in the IP-pool:
\begin{equation}\label{eq:cip}
c_{\rm IP}=\frac{n_{\rm IP}}{n_{\rm IP}^{\rm c}}-1,
\end{equation}
where $n_{\rm IP}^{\rm c}$ is the number of unique IPs present in the chunk averaged over all chunks, while $n_{\rm IP}$ has been defined in Sect.~\ref{sec:metrics}. This quantity takes the value $c_{\rm IP}=0$ when all the IPs are found in each chunk, i.e., when the IP-pool is stable and it is explored completely in each chunk (and therefore $n_{\rm IP}^{\rm c}=n_{\rm IP}$). 
On the other hand, if the IP-pool changes substantially from one chunk to the other, the total number of IPs $n_{\rm IP}$ is much larger than the average number of IPs $n_{\rm IP}^{\rm c}$ found in a chunk, and therefore $c_{\rm IP}$ becomes large (it is unbounded above). The same considerations apply to all the following metrics.

\emph{Change in the Set of Public Networks}.
While CDNs typically  use IPs taken from the same few public networks,  malicious fast flux frequently introduce IPs from new networks \cite{holz2008measuring,zhou2015survey}. We measure the change in the set of public networks by means of 
$c_{\rm net}=n_{\rm net}/n_{\rm net}^{c}-1$, where $n_{\rm net}^{\rm c}$ is the network-analogous of $n_{\rm IP}^{\rm c}$. 
 
\emph{Change in the Set of ASs}.
The generalisation of the previous argument to the next aggregation level brings us to the analysis of the changes in the number of AS involved. We introduce therefore $c_{\rm AS}=n_{\rm AS}/n_{\rm AS}^{c}-1$, where $n_{\rm AS}^{\rm c}$ is the AS-analogous of $n_{\rm IP}^{\rm c}$. 
 
\emph{Change in the Answer Length}. Another
relevant indicator is the change in the number of IPs retrieved in each query \cite{holz2008measuring,zhou2015survey}. We measure this change by means of $c_{\rm al}=m_{\rm al}/m_{\rm al}^{c}-1$, where $m_{\rm al}^{\rm c}$ is the $m_{\rm al}$-analogous of $n_{\rm IP}^{\rm c}$.

\subsection{Fast Flux Domains Identification}\label{sec:ff_identification}

A preliminary step for fast flux domains identification is the filtering of the queries with $d_{\rm IP}=0$, because this removes many false positives with no loss in terms of true positives.
The next step is the use of the metrics defined in Sect.~\ref{sec:metrics} to discriminate among malicious fast flux and CDN.
Instead of using a machine learning `black box' classifier, we combine the indicators in a controlled way, in order to encode some other domain knowledge and to allow for an easier interpretation of the results.
Foremost we aggregate the static and history-based metrics separately, and finally we combine them into a single anomaly indicator $A$, which can straightforwardly be used to classify the queries between fast flux and legit domains.

\subsubsection{Aggregation of the Static Metrics.}

We normalised the metrics $n_{\rm IP}$, $n_{\rm net}$, $n_{\rm AS}$, and $m_{\rm al}$ in the interval [0,1], so that for all of them the value 0 corresponds to a typical CDN, while 1 corresponds to the expected behaviour of a malicious fast flux. This is achieved by means of a square-exponential scaling of the form
\begin{equation}\label{eq:rescaling}
x\longrightarrow 1-e^{-\left(\frac{x-x_{0}}{s}\right)^2 },
\end{equation}
where $x_0=1$ is the minimum value for the metric before the rescaling, $s$ is different for each metric and represents an intermediate scale between a typical CDN behaviour and a behaviour clearly ascribed to a malicious fast flux.%
\footnote{The values of $s$ were set based on information retrieved from the literature (\cite{zhou2015survey} and references therein)
and the validation set. More in detail, we chose $s_{\rm IP}=24$, $s_{\rm net}=12$, $s_{\rm AS}=6$, and $s_{\rm al}=10$.}
Equation \ref{eq:rescaling} rescales $x=x_0$ (i.e., the smallest possible value for $x$), $x=s+x_0$ (i.e., a value   intermediate between the typical CDN behaviour and the typical malicious behaviour), and $x\gg s$ (i.e., a value much larger than the scale $s$) to $0$, $1/2$, and $1$ respectively.

After the scaling, all the quantities $n_{\rm IP}$, $n_{\rm net}$, $n_{\rm AS}$, $m_{\rm al}$, $f_{\rm AS}$, and $d_{\rm IP}$ are comparable: they take values in the interval [0,1] and for each of them a value close to 0 denotes a typical CDN behaviour, while a value close to 1 indicates a very suspicious behaviour. We combined these indicators with a weighted arithmetic mean in a unique static index%
\footnote{The weights reflect the importance of the corresponding metric in the correct classification in the validation set; the optimal values are $w_{\rm IP}=w_{\rm net}=0.03$, $w_{\rm AS}=0.13$, $w_{\rm al}=0.09$, $w_{f}=0.54$, and $w_{d}=0.18$.}
\begin{equation}\label{eq:stat}
A_{\rm stat}=w_{\rm IP}n_{\rm IP} + w_{\rm net}n_{\rm net} + w_{\rm AS}n_{\rm AS} + w_{\rm al}m_{\rm al} + w_{f} f_{\rm AS} + w_{d}d_{\rm IP}.
\end{equation}
In order to avoid the evaluation of misleading indicators due to lack of data, the metrics $f_{\rm AS}$ and $d_{\rm IP}$ are evaluated only if a minimum number of IPs is collected, while $m_{\rm al}$ is evaluated only if at least one answer contains more than one IP.
When one metric is absent, its value is set to 0 (in the absence of data we apply a sort of `presumption of innocence', to reduce false positives), its weight in the evaluation of $A_{\rm stat}$ is decreased by a factor 20 (because the innocence assessment is only due to the absence of data), and the other weights are proportionally rescaled so that $\sum_iw_i=1$.

\subsubsection{Aggregation of the History-Based Metrics.}

As already mentioned, the metrics $c_{\rm IP}$, $c_{\rm net}$, $c_{\rm AS}$, and $c_{\rm al}$ defined in Sect.~\ref{sec:metrics} are unbounded  above. We normalise them in the interval [0,1] by means of Eq.~\ref{eq:rescaling} with $x_0=0$ (as the minimum value for these metrics before the rescaling  is 0).%
\footnote{The values of $s$ were set based on information retrieved from the literature  and the validation set. More in detail, we chose $s_{\rm IP}=s_{\rm net}=1$ and $s_{\rm AS}=s_{\rm al}=0.5$. }
After the rescaling, all the metrics take values in the interval [0,1] and for each of them a value close to 0 corresponds to a very stable behaviour, while a value close to 1 indicates a behaviour with high variability over time. 
We combine then in a unique indicator three of the history-based metrics (the fourth, i.e., $c_{\rm al}$ is instead used in Eq.~\ref{eq:A}) with a weighted arithmetic mean%
\footnote{The weights reflect the importance of the corresponding metric in the validation set; the optimal values are $w'_{\rm IP}=0.07$, $w'_{\rm net}=0.23$, and $w'_{\rm AS}=0.7$.}
\begin{equation}\label{eq:dyn}
A_{\rm dyn}=w'_{\rm IP}c_{\rm IP} + w'_{\rm net}c_{\rm net} + w'_{\rm AS}c_{\rm AS}.
\end{equation}

\subsubsection{Final Aggregation.}

We combine the indicators $A_{\rm stat}$, $A_{\rm dyn}$, and $c_{\rm al}$ into a single anomaly indicator $A$, which should be used to classify the queries between fast flux and legit domains.
In order to reduce false positives, we differentiate on the basis of the quantity $f_{\rm AS}$, and we define
\begin{equation}\label{eq:A}
 A=\left\{\begin{array}{lcl}
  \sum_i w_iA_i & {\ \rm if\ } & f_{\rm AS}\geq 0.5\\
  \prod_i (A_i)^{w_i} & {\ \rm if\ } & f_{\rm AS}<0.5\\
 \end{array}   \right. ,
\end{equation}
 where $\{ A_i\}=\{A_{\rm stat}, A_{\rm dyn},c_{\rm al}\}$ and $\{w_i\}$ are the related weights.%
\footnote{An optimisation procedure on the validation set produced similar weight for the three quantities: $w_{\rm stat}=0.27$, $w_{\rm  dyn}=0.38$, and $w_{\rm al}=0.35$.}
Analogously to the averages in Eqs.~\ref{eq:stat} and \ref{eq:dyn}, when one metric is absent, its value $A_i$ is set to 0 (not anomalous), its weigth $w_i$ is decreased by a factor 20, and the other weights are proportionally rescaled so that $\sum_jw_j=1$.

Note that in Eq.~\ref{eq:A} a (weighted) arithmetic mean is used when the AS-fraction is large, while a (weighted) geometric mean is used when the AS-fraction is small; this implies that in the latter case a value close to 0 for one of the indicators $A_i$ gives a stronger penalty to $A$.

The detection of malicious fast flux has thus been reduced to a very simple one-dimension classification problem: only queries with $A>A_{\rm th}$ are labeled as fast flux, where the optimal threshold  ($A_{\rm th}=0.25$) has been found  by maximizing the performance on the validation set. 
In order to increase the readability of the results, we applied a sigmoid-shaped rescaling which maps $A=0$ and $A=1$ onto themselves and $A_{\rm th}$ onto $0.5$.

\section{Experimental Evaluation}\label{sec:experimental}

The  fast flux detection algorithm described in Sect.~\ref{sec:detection} was evaluated over a test set comprising 30 days of ordinary traffic of the network described in Table~\ref{tab:network} with the injection of fast flux traffic which covers all the most relevant fast flux attack scenarios (see Table~\ref{tab:malware} for a complete list).

%
%
\begin{table}[!ht]
\centering
\caption{Test-network description}\label{tab:network}
\advance\leftskip-2mm
\begin{tabular}{ | p{7.1cm} | c | c |}
  \hline
  &\,\textbf{30-days total}\,&\,\textbf{one-hour average}\,\\
  \hline
   \textbf{N. of machines} & 391 & - \\
  \hline
  \textbf{N. of client machines} & 286 & - \\
  \hline  
  \textbf{N. of connections} & 398\,M & 
  552\,k \\
  \hline
  \textbf{N. of resolved A-type DNS queries} & 75\,M  & 104\,k \\
  \hline
  \textbf{N. of unique resolved A-type DNS queries} & 1.3\,M & 1.9\,k\\
  \hline
\end{tabular}
\end{table}
%
%

The fast flux traffic has been injected in the network via 47 pcaps --- collected from the public repositories \cite{sandbox1,sandbox3,sandbox2} --- which are associated with 9 different malware campaigns.
Table \ref{tab:malware} provides a brief description of each malware campaign with the following information:
\begin{itemize}
 \item the category, i.e., the malware type associated with the campaign
 \item the name of the campaign
 \item the list of the domains present in each pcap of the campaign, with the anomaly indicator $A$ associated by the algorithm to each of them
 \item the average value of $A$ for each campaign
\end{itemize}
In order to rule out overfitting, we used the test set (i.e., the network traffic described in Table \ref{tab:network} and the pcaps described in Table \ref{tab:malware}) only to test the performance of the algorithm, while we used another set (see Sect.~\ref{sec:detection}) to define the algorithm and the values of its parameters.

%
%
\begin{table*}[!ht]
\centering
\advance\leftskip-5mm
\begin{minipage}{\textwidth}
\caption{Malware description (the underlined domains are farther analysed in Sect.~\ref{sec:botnet})}\label{tab:malware}
\begin{tabular}{ | C{1.8cm} | C{1.84cm} | C{8cm} | c |}
  \hline
  \textbf{Category} & \textbf{Campaign} & \textbf{Domains ($A$)} & \textbf{$\langle A\rangle$} \\
  \hline
Banking Trojan & ZBOT & miscapoerasun.ws (0.85) & 0.85 \\
\hline
Banking Trojan & Dreambot & rahmatulahh.at (0.89); ardshinbank.at (0.92) &\ 0.91\ \\
\hline
Banking Trojan & Ursnif & widmwdndghdk.com (0.90); bnvmcnjghkeht.com (0.85); qqweerr.com (0.85) & 0.87 \\
\hline
VBA Dropper & Doc Dropper Agent\footnote{Doc.Dropper.Agent-6332127-0 \cite{vbadropper}} & aassmcncnnc.com (0.90); iiieeejrjrjr.com (0.87); ghmchdkenee.com (0.88) & 0.88 \\
\hline
Ransomware & Locky & thedarkpvp.net (0.83); nsaflow.info (0.91); mrscrowe.net (0.93); sherylbro.net (0.87); \underline{scottfranch.org} (0.90); gdiscoun.org (0.90)  & 0.89 \\
\hline
Ransomware & Nymaim & iqbppddvjq.com (0.91); danrnysvp.com (0.91); pmjpdwys.com (0.93); vqmfcxo.com (0.86); gbfeiseis.com (0.91);  \underline{iuzngzhl.com} (0.97); \underline{vpvqskazjvco.com} (0.84); \underline{jauudedqnm.com} (0.93); \underline{dtybgsb.com} (0.93); \underline{tuzhohg.com} (0.93); \underline{sxrhysqdpx.com} (0.86); \underline{arlfbqcc.com} (0.93); danrnysvp.com (0.87) & 0.90 \\
\hline
Banking Trojan & Zeus Panda & \underline{farvictor.co} (0.89); \underline{fardunkan.co} (0.89); \underline{bozem.co} (0.84); \underline{farmacyan.co} (0.87); \underline{fargugo.co} (0.90); \underline{manfam.co} (0.85) & 0.87 \\
\hline
Banking Trojan & GOZI ISFB & \underline{qdkngijbqnwehiqwrbzudwe.com} (0.80); \underline{jnossidjfnweqrfew.com} (0.90); \underline{zxciuniqhweizsds.com} (0.86); huwikacjajsneqwe.com (0.92); efoijowufjaowudawd.com (0.92); onlyplacesattributionthe.net (0.90); nvvnfjvnfjcdnj.net (0.86); popoiuiuntnt.net (0.89); zzzzmmmsnsns.net (0.80); popooosneneee.net (0.83); liceindividualshall.net (0.87); roborobonsnsnn.net (0.93) & 0.87 \\
\hline
Ransomware & GandCrab & \underline{zonealarm.bit} (0.90) & 0.90 \\
  \hline  
\end{tabular}
\end{minipage}
\end{table*}
%
%

Table \ref{tab:malware} clearly shows that  the proposed method successfully detected all the fast flux domains with a high anomaly indicator. In fact the value of $A$ averaged over all campaigns is equal to 0.89.

%
%
\begin{table}[!ht]
\caption{Results}\label{tab:results}
\centering	
\begin{tabular}{ | c | c | c |}
  \hline
  & \textbf{$A>0$} & \textbf{$A>0.5$} \\
  \hline
  \textbf{True Positives ($T_P$)} & 47 (100\%) & 47 (100\%) \\
  \textbf{\ False Negatives ($F_N$)\ } & 0 (0\%) & 0 (0\%) \\
  \textbf{False Positives ($F_P$)} & 6 ($<$0.001\%) & 4 ($<$0.001\%) \\
  \hline
\end{tabular}
\end{table}
%
%

In Table~\ref{tab:results} we summarise the performance of the algorithm: in the second column we consider the total number of outputs of the algorithm (i.e., the number of domains with $A>0$) while in the third column we report the number of outputs labeled as fast flux (i.e., the number of domains with $A>0.5$).
On the rows we report the following quantities
\begin{itemize}
 \item True Positives rate ($T_P$): the number of unique fast flux domains detected;
 \item False Negatives rate ($F_N$): the number of unique fast flux domains incorrectly labeled as legit;
 \item False Positives rate ($F_P$): the number of unique legit domains incorrectly labeled as fast flux.
\end{itemize}
 All rates are given as absolute values and as percentages for each type on the corresponding number 
 of unique domains in input.

A remarkable result is the absence of false negatives: this determines indeed a 100\% recall, also known as detection rate, $R=T_P/(T_P+F_N)$.
In order to evaluate the algorithm also with a metric that takes into account the false positives rate $F_P$, we computed the F-score $F=2\,P\,R/(P+R)$ (where $P=T_P/(T_P+F_P)$), obtaining $F=95.9\%$. 

As a comparison, \cite{soltanaghaei2015detection} obtained $R=94.4\%$ and $F=89.5\%$
in their best experimental result. We can therefore conclude that the proposed method is able to detect queries to fast flux domains in a corporate network in near-real-time and with high anomaly indicators, limiting false positives at the same time.

\section{Fast Flux Service Networks Analysis}\label{sec:botnet}

As an in-depth analysis of the algorithm described in Sect.~\ref{sec:detection}, we examined the IPs  associated to a list of fast flux domains. 
The IPs were collected via active DNS analysis and precisely with an FFSN-spanner which resolved systematically domains taken from a list of malicious domains; 
these domains were gathered via a scouting activity from the public repositories \cite{sandbox1,sandbox3,sandbox2,virustotal}. 
With the purpose of hiding the FFSN-spanner activity from the bot herders, we randomized the sequence of the queries and the waiting times among two subsequent queries, while implementing an anonymization technique based on the use of the Tor network. In order to overcome a limitation of the DNSPort resolver \cite{tor}, which returns only the first answer for domain lookup, we adopted \emph{ttdnsd}, the Tor TCP DNS Daemon.
This solution allows for making arbitrary DNS requests  by converting any UDP request into a TCP connection, which is given to Tor through the SOCKS port. 
The request is then forwarded anonymously through the Tor network and reaches one of the `open' recursive name servers via the Tor Exit node.

\begin{figure}[ht]
\centering
\includegraphics[width=11.5cm]{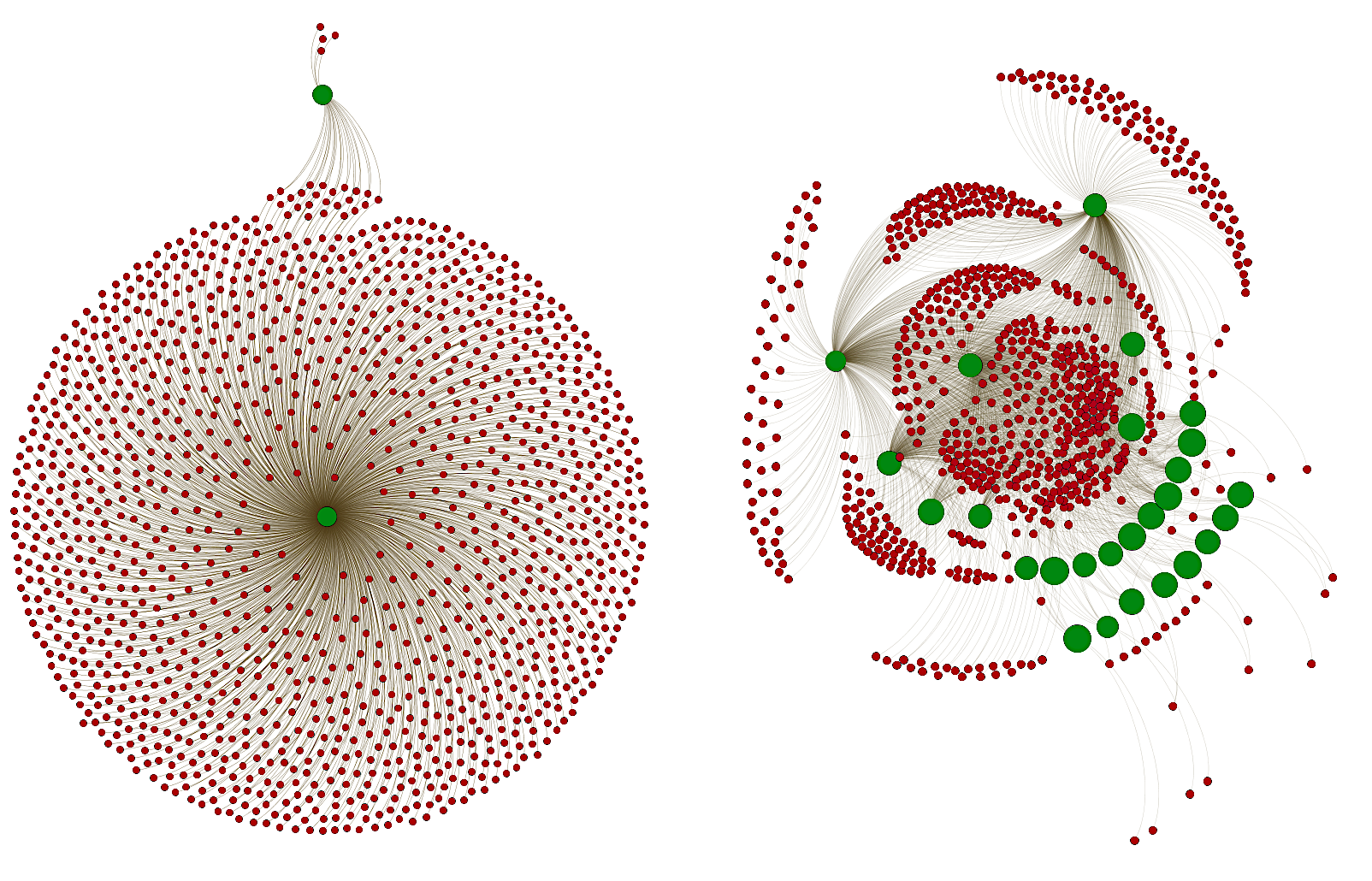}
\caption{\label{fig:botnet-graph} Bipartite graph representing the IPs (\textit{small red circles}) associated with each domain (\textit{large green circle}) in the pre-migration scenario. An arc 
indicates that the IP has been given in answer to a query resolving the domain.}
\end{figure}
%

\begin{figure}[ht]
\begin{center}
\advance\leftskip-8mm
\includegraphics[width=12cm]{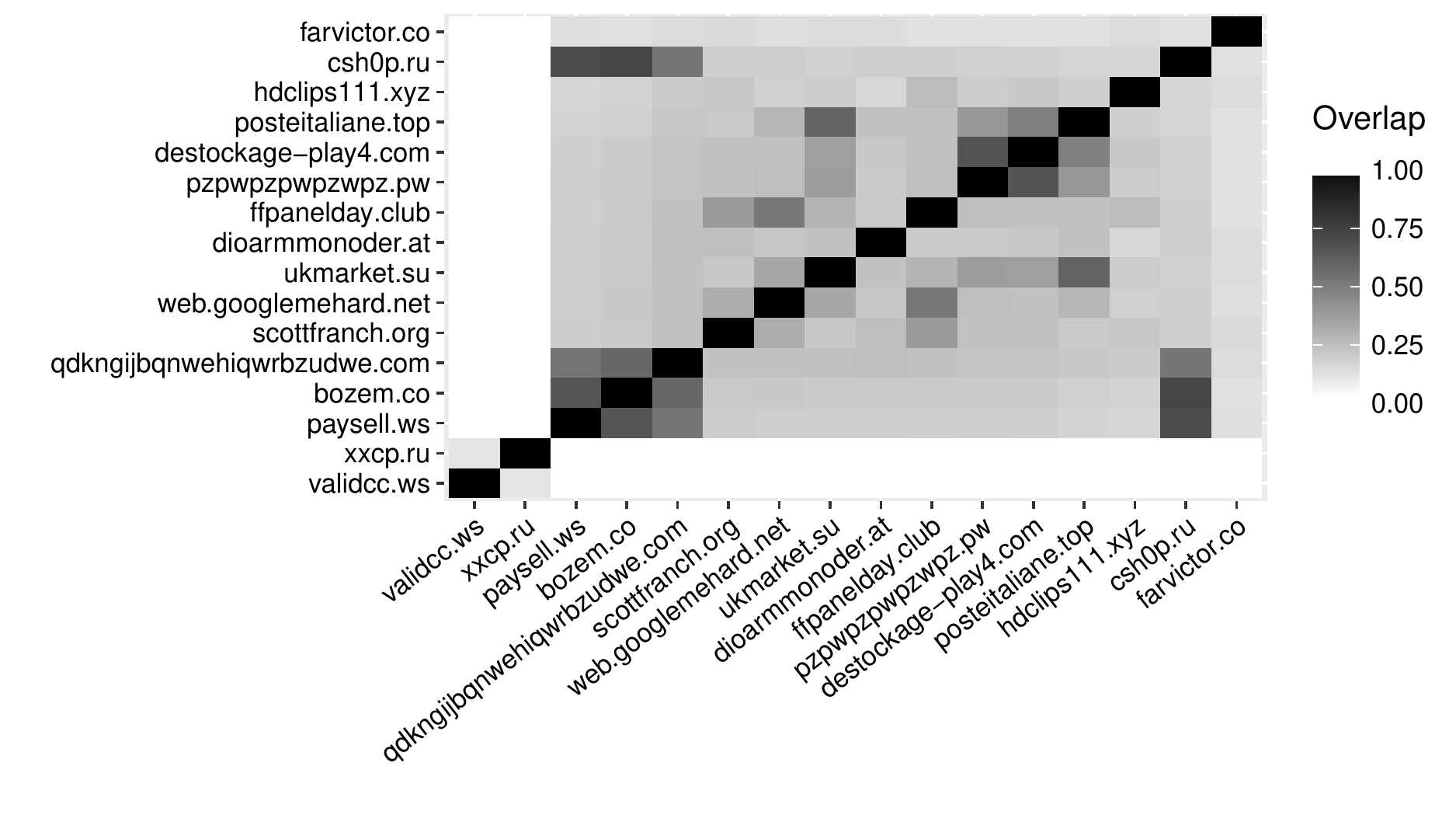} 
\caption{\label{fig:IP_overlap} 
Overlap representation $O_{ij}$ (defined in Eq.~\ref{eq:overlap}) among all the pairs $(i,j)$ of the top 16 domains (for number of retrieved IPs) in the pre-migration scenario. Darker tones represent larger overlaps.}
\end{center} 
\end{figure}

Over the period 09.03.2018--19.04.2018, we collected 10747 IPs associated with 55  domains%
\footnote{Some domains are reported in Table \ref{tab:malware}, others in Fig.~\ref{fig:IP_overlap}; the remaining domains are odqndpqowdnqwpodn.com, moncompte-carrefour.org, 0768.ru, allianzbank.org, commerzb.co, db-ag.co, druhok.com, form.xbeginner.org, ihalbom.com, ingdirectverifica.com, lloyds-personal.com, mein-advanzia.info, point.charitablex.org, postofficegreat.com, ransomware.bit, redluck0.com, safe.bintrust.org, sunyst.co,  dfplajngru.com, mer.arintrueed.org, www.ico-teleqram.net, clo.arotamarid.org, www.translationdoor.com, vr-b.co, vr-b.cc.}%
: 7640 are fictitious IPs, related to the Nymaim campaign, while the remaining 3017 IPs are likely associated with compromised hosts. The IPs of the first group (\emph{Nymaim-fake-IPs}), which are translated into real IPs%
\footnote{An analysis on some pcaps associated with iuzngzhl.com, arlfbqcc.com, and vpvqskazjvco.com revealed that the corresponding real IPs are based on the SandiFlux FFSN described below.}
by the decoder algorithm in the malware that use them \cite{nymaim}, are strictly related to the \emph{C\&C Network} described in \cite{katz2017digging}.
The IPs of the latter group (\emph{real IPs}) showed instead a relevant change in the behaviour on the 26.03.2018: this was probably the last part of the migration described in \cite{sandiflux}.

The pre-migration scenario is described in Fig.~\ref{fig:botnet-graph}: it can be noted that different domains (\textit{large green circles}) share some IPs (\textit{small red circles}).
In Fig.~\ref{fig:IP_overlap} we represent the overlap $O_{ij}$ among all the pairs $(i,j)$ of the top 16 domains observed before the migration (excluded the ones associated with the Nymain campaign),
defined as
\begin{equation}\label{eq:overlap}
 O_{ij}=\frac{\left|X_i\cap X_j\right|}{|X_i\cup X_j|},
\end{equation}
where $X_i$ is the pool of IPs associated with the $i$-th domain and $|X|$ is the the cardinality of $X$.

Both Figs.~\ref{fig:botnet-graph} and \ref{fig:IP_overlap} show a clear subdivision of the domains in two independent clusters.
The analysis of the fast flux domains revealed that the clusters correspond to two large FFSNs, namely Dark Cloud (on the left in Figs.~\ref{fig:botnet-graph} and \ref{fig:IP_overlap}) and SandiFlux (on the right).
Indeed, in the first cluster we recognised domains associated with  Dumps Stores that leverage Dark Cloud \cite{crowder2016dark}, while in the latter we found domains associated with the GandCrab campaigns, which leverage SandiFlux \cite{sandiflux}.
It is worth noting that the sets of IPs in the two FFSNs that we identified are highly overlapped, but no IP is shared among the two groups.

\begin{figure}[ht]
\begin{center} 
\includegraphics[width=12cm]{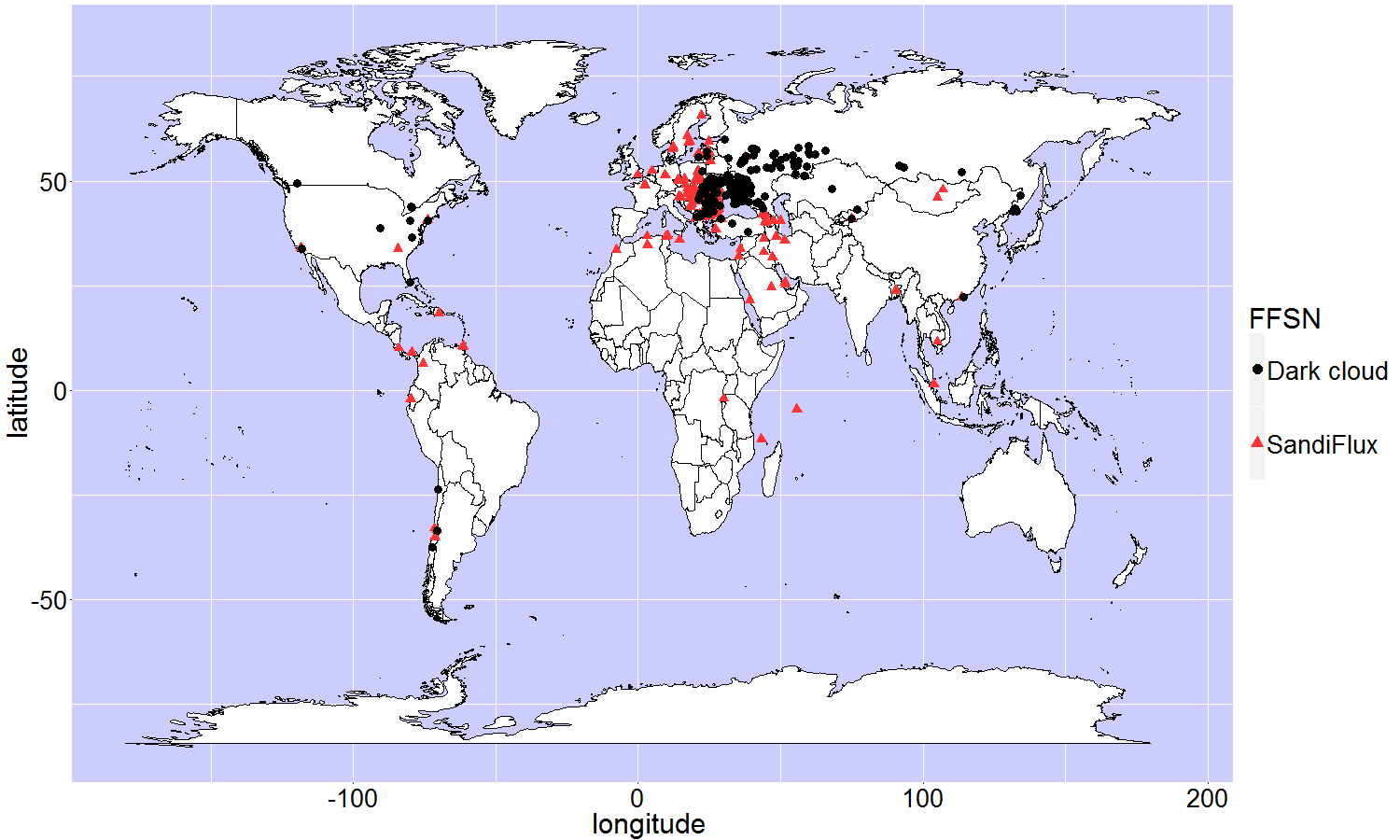} 
\caption{\label{fig:geoloc} Geolocation of the IPs retrieved for the FFSNs before the migration} 
\end{center} 
\end{figure} 
\begin{figure}[ht]
\begin{center} 
\includegraphics[width=12cm]{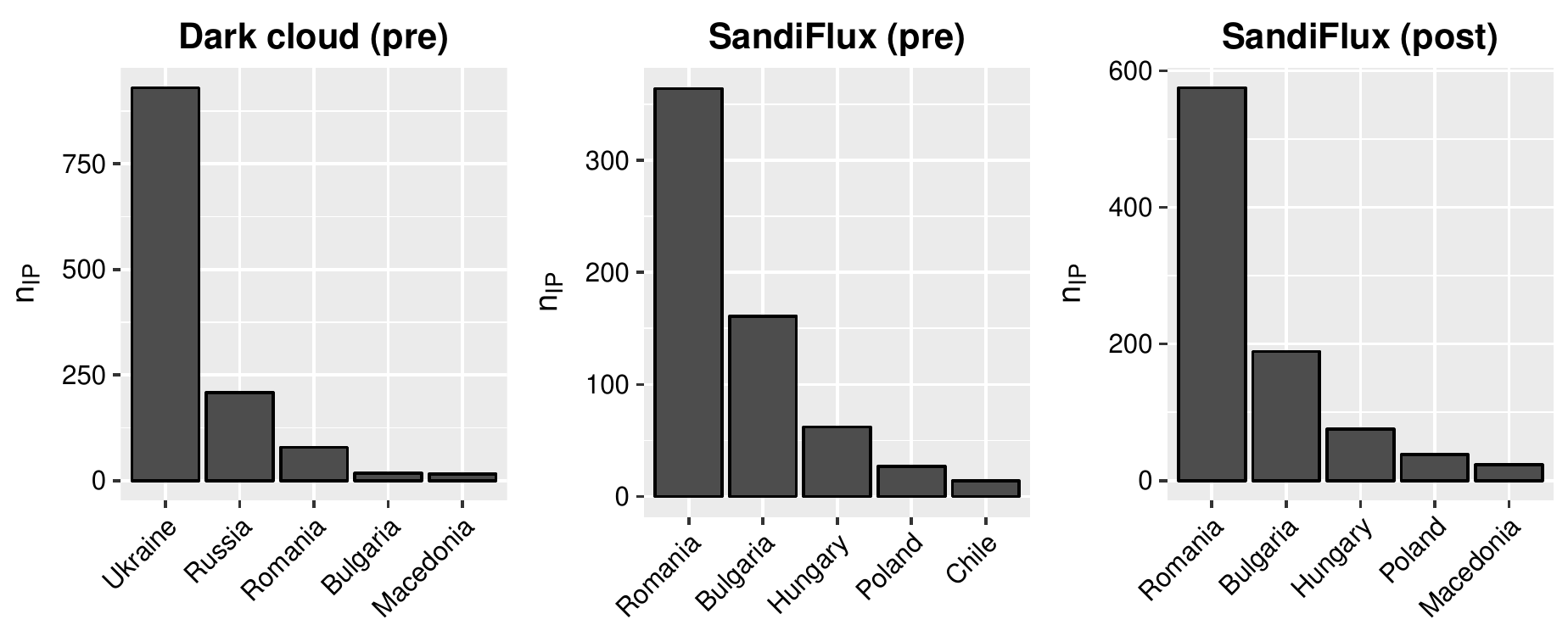} 
\caption{\label{fig:bycountry} Histogram of the number of IPs localised in the top 5 countries  before the migration (on the left) and after (on the right)}. 
\end{center} 
\end{figure}

The pre-migration subdivision in two different FFSNs is reflected in the different geolocation of the relative IPs: Fig.~\ref{fig:geoloc} shows that, while the IPs retrieved from SandiFlux are mainly localised in central-east Europe, the ones retrieved from Dark Cloud are based in eastern Europe and Russia.

After the migration all the domains appeared to leverage a unique FFSN, which we recognised as SandiFlux.
In Fig.~\ref{fig:bycountry} we further investigate the top 5 countries represented before the migration (on the left) and after (on the right): Ukraine and Russia confirmed to be the most represented countries in Dark Cloud \cite{crowder2016dark}, while SandiFlux's IPs are found mainly in Romania and Bulgaria both before and after the migration.
Figures \ref{fig:geoloc} and \ref{fig:bycountry} are based on the IP-geoloc tables downloaded from Maxmind \cite{maxmind}.

\begin{figure}[ht]
\begin{center} 
\includegraphics[width=12cm]{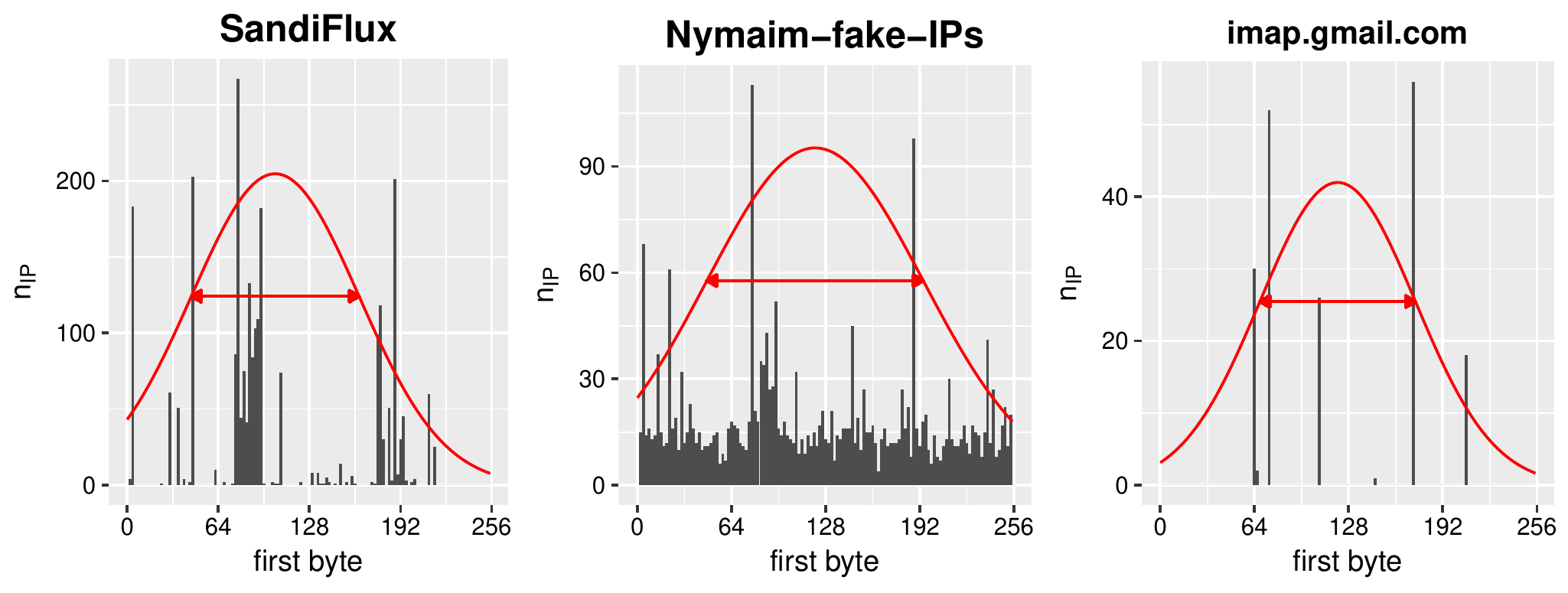} 
\caption{\label{fig:ip_distribution} Histogram of the frequencies of the first byte in the IP-pool associated with two groups of fast flux domains and with one large CDN (bin-size=2)} 
\end{center} 
\end{figure}

The FFSNs described above are a good testing ground for the metrics introduced in Sect.~\ref{sec:metrics}: 
in Table~\ref{tab:metrics}  we report a summary of some of these metrics evaluated over the FFSNs and three large CDNs. Note  that two of the CDNs we observed (namely, www.nationalgeographic.it and cdn.wetransfer.net) have a very large number of IPs, making thus $n_{\rm IP}$ a misleading indicator in these cases.
This is not a problem for the proposed algorithm, since, as explained in Sect.~\ref{sec:detection}, $n_{\rm IP}$ is used in combination with many other metrics.
%
%
\begin{table}[!ht]
\caption{Summary of some relevant metrics}\label{tab:metrics}
\centering	
\begin{tabular}{ |p{3.9cm}|c|c|c|c|c|c|c| }
\hline

&\ $n_{\rm IP}$\ &\  $n_{\rm AS}$\ &\  $n_{\rm IP}^{\rm resc}$\ &\  $n_{\rm AS}^{\rm resc}$\ &\ $c_{\rm AS}^{\rm resc}$\ &\ $f_{\rm AS}^{\rm resc}$\ &\  $d_{\rm IP}$\   \\
\hline

\textbf{Dark Cloud} & \ 1276 \ & 221 & 1 & 1 & 1 & 1 & $1.1\ 10^{-3}$ \\

\textbf{SandiFlux} & \ 1831 \ & 354 & 1 & 1 & 1 & 1 & $1.3\ 10^{-3}$ \\

\textbf{Nymaim-fake-IPs} &  7640 &\ 1767 \ & 1 & 1 &  1 & 1 & 0.77 \\

www.nationalgeographic.it  & 2589 & 1 & 1 & 0 & 0 & 0 & \ $2.8\ 10^{-6}$ \     \\ 

cdn.wetransfer.net & 2852 & 1 & 1 & 0 & 0 & 0 & $3.1\ 10^{-6}$  \\

neo4j.com & 33 & 1 & 0.83 & 0 & 0 & 0 & $5.8\ 10^{-4}$  \\

\hline
\end{tabular}
\end{table}
%
%

In Fig.~\ref{fig:ip_distribution} we represent the histogram of the frequencies of the first byte in the IP-pool of SandiFlux and Nymaim-fake-IPs and one medium-size CDN. 
A clear difference between botnets can be noticed, in particular among the Nymaim-fake-IPs, where the IP-distribution is not so far from a uniform random distribution and the CDN `imap.gmail.com', where the IP-distribution has a few high peaks.
Figure \ref{fig:ip_distribution} clearly shows that simple indicators as the mean and the variance (represented by the corresponding Gaussian distribution) 
do not catch the nature of the distribution, while the metric $d_{\rm IP}$ defined in Eq.~\ref{eq:ipdispesion} is much more appropriate. In particular 
$d_{\rm IP}=1.3\ 10^{-3}$ for SandiFlux
 and $d_{\rm IP}=0.77$ for Nymaim-fake-IPS, while the CDN `imap.gmail.com' has $d_{\rm IP}=5.0\ 10^{-8}$.
Note that the encoding of the IPs is not a problem for the detection algorithm: in fact it increases the IP dispersion, thus fastening the detection.

\section{Conclusions}\label{sec:conclusions}

In this paper, we proposed a  fast flux detection method based on the passive analysis of the DNS traffic of a corporate network. The analysis is based on \emph{aramis} security monitoring system. 
The proposed solution has been evaluated over the LAN of a  company, with the injection of 47 pcaps associated with 9 different malware campaigns that leverage FFSNs and cover a wide variety of attack scenarios. All the fast flux domains were detected with a very low false positive rate and the comparison of performance indicators with a state-of-the-art work shows a remarkable improvement.
An in-depth active analysis of a list of malicious fast flux domains confirmed the reliability of the metrics used in the proposed algorithm and allowed for the identification of more than 10000 IPs, some of which are likely associated with compromised hosts. These IPs turned out to be part of two notorious botnets, namely Dark Cloud and SandiFlux, to the description of which we therefore contribute.

As a future development, we plan to introduce in the algorithm a metric related to the use of reserved IPs, which we observed to be extensively present in SandiFlux.
Another planned development is the inspection of the overlap in terms of IPs among the most suspicious domains, as we saw that many IPs are shared among domains in the same FFSN.


%
%
\bibliographystyle{splncs03} 
\bibliography{references}

%
%
%
%
%
%
%
%
\end{document}